\documentclass[conference]{IEEEtran}
\IEEEoverridecommandlockouts
\usepackage{cite}
\usepackage{amsmath,amssymb,amsfonts}
\usepackage{algorithmic}
\usepackage{graphicx}
\usepackage{textcomp}
\usepackage{slashbox}
\usepackage{xcolor}

\begin{document}

\title{
Loneliness Forecasting Using Multi-modal Wearable and Mobile Sensing in Everyday Settings}

\author{\IEEEauthorblockN{ Zhongqi Yang$^1$,
Iman Azimi$^1$,
Salar Jafarlou$^1$,
Sina Labbaf$^1$,
Brenda Nguyen$^2$,
Hana Qureshi$^2$,\\
Christopher Marcotullio$^2$,
Jessica L. Borelli$^2$,
Nikil Dutt$^1$,
and Amir M. Rahmani$^{1,3}$}
\IEEEauthorblockA{\textit{$^1$Department of Computer Science}\\
\textit{$^2$Department of Psychological Science}\\\textit{$^3$School of Nursing}\\
\textit{University of California, Irvine}\\
\{zhongqy4, azimii, jafarlos, slabbaf, brendn3, hanaq, cmarcotu, jessica.borelli, dutt, a.rahmani\}@uci.edu}}

\maketitle

\footnote{Accepted to IEEE-EMBS International Conference on
Body Sensor Networks:
Sensor and Systems for Digital Health
(IEEE BSN 2023)
}
\begin{abstract}
The adverse effects of loneliness on both physical and mental well-being are profound. Although previous research has utilized mobile sensing techniques to detect mental health issues, 
few studies have utilized state-of-the-art wearable devices to forecast loneliness and estimate the physiological manifestations of loneliness and its predictive nature.
The primary objective of this study is to examine the feasibility of forecasting loneliness by employing wearable devices, such as smart rings and watches, to monitor early physiological indicators of loneliness. Furthermore, smartphones are employed to capture initial behavioral signs of loneliness.
To accomplish this, we employed personalized machine learning techniques, leveraging a comprehensive dataset comprising physiological and behavioral information obtained during our study involving the monitoring of college students. Through the development of personalized models, we achieved a notable accuracy of 0.82 and an F-1 score of 0.82 in forecasting loneliness levels seven days in advance. Additionally, the application of Shapley values facilitated model explainability.
The wealth of data provided by this study, coupled with the forecasting methodology employed, possesses the potential to augment interventions and facilitate the early identification of loneliness within populations at risk.
\end{abstract}

\section{Introduction}

Loneliness is a negative feeling often related to loss and disappointment~\cite{perlman1981toward}. 
It arises when individuals evaluate their existing relationships against their own wishes and societal expectations~\cite{perlman1981personal}. 
Loneliness can have negative impacts on physical and mental health, leading to heightened rates of morbidity and mortality~\cite{holt2015loneliness}.
Multiple studies, such as the research conducted by Park \textit{et al.}~\cite{park2020effect}, have provided compelling evidence establishing a connection between loneliness and adverse physiological and mental health outcomes. Notably, associations have been identified between loneliness and disturbances in sleep patterns, as well as diminished cardiac output.
Furthermore, the ongoing COVID-19 pandemic has amplified concerns surrounding loneliness, particularly among adolescents and young adults hailing from lower socioeconomic backgrounds ~\cite{banerjee2020social}. The profound implications for health underscore the imperative of comprehending the underlying factors contributing to the development of loneliness, as well as the optimal timing for its detection and forecasting.

Existing studies have capitalized on wearable and mobile sensing to detect loneliness.
Such works encompass the use of smartphones, wearable sensors, wearable activity trackers, and other technologies to collect information on environmental context, heart rate, activity, and sleep patterns.
For instance, Wu \textit{et al.} utilize smartphones to collect geosocial data such as the individual's location and social interaction to detect loneliness in real-time~\cite{wu2021improving}. 
Li \textit{et al.} capture the app usage information as additional features to identify loneliness risk~\cite{li2016loneliness}.
Doryab \textit{et al.} employ smartphones to capture behavioral data, including locations, calls log, screen status, etc. as well the FitBit sensors to capture mobility and physical activity to detect loneliness~\cite{doryab2019identifying}. 
% As a sloe smartphone can only provide limited information based on the frequency of smartphone usage and may not fully capture the nuances of loneliness experiences. 
% Several recent studies have attempted to incorporate passive sensors due to their advantage of enabling passive monitoring, which eliminates the need for active user engagement.
Besides loneliness, a few studies incorporate mobile sensing to identify other mental health outcomes such as stress, depression, and suicidal ideation~\cite{sano2018identifying,haines2020testing,narziev2020stdd}. 

% \textbf{Divided into 1.loneliness 2. other similar to loneliness (in addition to loneliness.....)}

% These studies focus only on prompt loneliness detection.... they do not investigate the variation on time series information ffrom the objective data/ relation between these and loneliness

% are not able to \textbf{forecast}

% \textbf{Sensors are not problem}

% However, the previous passive sensor-based studies are limited by the capability of sensors to collect various signals.
% The potential of cutting-edge wearable devices for such tasks remains largely unexplored.

% These studies focus only on prompt loneliness detection.... they do not investigate the variation on time series information ffrom the objective data/ relation between these and loneliness

% The studies on mental health status detection, especially loneliness, predominantly concentrate on the detection at the present moment or in real-time.
% There is a gap when it comes to forecasting loneliness with a time gap between the predictive features and the target outcome.
% Specifically, existing studies have not explored the prediction of loneliness levels with a time lag, i.e., forecasting loneliness after a certain period. 
% Due to the absence of studies on forecasting loneliness, there is currently no dedicated dataset available for this specific research question.

Nevertheless, current studies predominantly concentrate on the immediate detection of loneliness or offline analysis, failing to account for temporal variations in time series data or explore the interplay between objective data and loneliness. This glaring gap in the literature pertains specifically to the forecasting of loneliness levels with a time interval between the predictive features and the target outcome.
More precisely, there exists a lack of research pertaining to the forecasting of loneliness over a defined period, leading to an insufficient availability of dedicated datasets for addressing this specific research inquiry.

% Nevertheless, there still lacks study  leveraging cutting-edge wearable devices on loneliness detection. 

% \textbf{In the following 3 paragraphs, I wanna say that these devices, and collecting a dedicated dataset, are promising on this task.}
Modern wearable devices (such as the Oura Ring and Samsung smartwatch) and mobile phones have demonstrated their capacity to assess a wide range of physiological, behavioral  and contextual information. Their viability for integration into healthcare applications has been substantiated by previous research~\cite{nissen2022heart,wang2014studentlife}.
The incorporation of such supplementary information holds promise as a prospective avenue for exploration in the modeling of loneliness.
% Also, to the best of our knowledge, there are no studies on loneliness forecasting with a time gap, and there is no such dataset that is dedicated to the specific research question. 
% The existing literature focus on loneliness modeling with historical data or detection in real-time. 
Furthermore, with regard to the intricate relationship between loneliness and other health risks, the potential forecasting or early detection of feelings of loneliness presents novel opportunities for the development of targeted interventions aimed at alleviating loneliness and enhancing individual well-being.
By leveraging the ubiquitous sensing capabilities of modernized devices, it becomes possible to continuously gather an extensive array of physiological and behavioral data in tandem with self-reported levels of loneliness to build models.

% By leveraging the comprehensive dataset with additional information, particularly the unique factors associated with loneliness, we can expand our understanding of the underlying mechanisms and develop forecasting models for loneliness.

% This data includes sleep-related features such as duration, quality, and restlessness, as well as cardiovascular indicators such as heart rate, heart rate variability, and blood pressure~\cite{}.

% we can build upon the existing literature and achieve a more comprehensive and detailed understanding of the occurrence and mechanisms of loneliness. 
% Moreover, the collection of such comprehensive data offers opportunities that extend beyond a mere understanding of the mechanisms and associations between loneliness and health outcomes. 
% By leveraging this data, it becomes feasible to predict loneliness with satisfactory accuracy and granularity.

In this paper, we present a personalized machine-learning method to forecast levels of loneliness by leveraging longitudinal data encompassing physiological, behavioral, and contextual information collected from smart rings, watches, and smartphones. To achieve this, we conducted a two-month monitoring study involving a cohort of 29 college students, utilizing the Oura Ring, Samsung Smartwatch, and the AWARE app installed on smartphones.
In addition to capturing objective and passive data through these devices, we collected self-reported loneliness information from the participants, thereby generating detailed and granular labels for evaluation purposes.
The collected objective and subjective data were employed to train and assess the effectiveness of our proposed loneliness forecasting method. Moreover, we integrated SHAP (SHapley Additive exPlanations) to compute the Shapley values, allowing for a deeper understanding of the relationship between the temporal features derived from these time series data and the experience of loneliness~\cite{lundberg2017unified}.
%This represents an advancement beyond existing studies that primarily focus on understanding the mechanisms of loneliness, as the forecasting has the potential for further clinical implementation as a means of early identifying and addressing loneliness among at-risk populations.
%To facilitate this research, we have curated a comprehensive dataset that includes physiological data from wearable devices and behavioral information from AWARE. 
%The utilization of this rich dataset in future research can greatly contribute to the development of effective interventions for loneliness.

% Secondly, we propose a machine leanring method to forecast loneliness categories based on physiological time series collected from ubiquitous wearable devices such as Oura Ring and Samsung Smartwatch, as well as behavior feature extraction from smartphone usage information collected from AWARE.
% We also implement explainability methods to find insight into the relationship between these time series and loneliness.
% This represents an advancement beyond existing studies that primarily focus on understanding the mechanisms of loneliness, as the forecasting has the potential for further clinical implementation as a means of early identifying and addressing loneliness among at-risk populations.

\section{Loneliness Data Collection}
% Perform data collection/methods to collect data

Our study employs a comprehensive approach to collecting data, including self-reported loneliness and related physiological and behavioral data. We utilized the ZotCare platform for efficient data collection and storage~\cite{labbaf2023zotcare}.
The gathered data can be categorized into three types: objective physiological, objective behavioral, and subjective questionnaires. Objective physiological data were collected using the Oura ring and Samsung smartwatch.
The Oura ring assessed sleep patterns, while the Samsung smartwatch continuously monitored physiological parameters throughout the day.
Objective behavioral data were collected through the implementation of the AWARE~\cite{ferreira2015aware} phone app, which captured mobile sensory data and recorded detailed information regarding participants' phone usage patterns.
Subjective self-report data were gathered employing our custom mSavorUs~\cite{msavor} phone app, specifically designed for the collection of subjective experiential data and the provision of interventions. The self-reported loneliness levels were measured on a scale ranging from 0 to 100, with 0 indicating ``not feeling lonely at all" and 100 indicating ``feeling extremely lonely." These self-reported loneliness levels were further categorized as either "Not Feel Lonely" or "Feel Lonely," based on whether they fell below or above the population median, respectively. All procedures were approved by the researchers’ academic institutional review board (\#2019-5153).

\subsection{Participants}

This study explores loneliness and mental health in 29 full-time college students from a Southern California university. Excluded are married students, those with children, returnees after a three-year hiatus, and severe psychopathology cases.
Participants needed fluent English and an Android smartphone compatible with Oura Ring and Samsung Active 2 watch for data collection. Recruitment involved faculty outreach, social media sharing, and screening surveys for depression/suicidal ideation.

% Participants meeting the criteria received support from our clinical psychologist and access to campus wellness resources.
\subsection{Recruitment}
% Participants meeting the inclusion criteria underwent a baseline assessment during an in-person lab session. 
% They set up and downloaded the Oura Ring, Samsung Gear Sport watch, and study-related smartphone apps under instruction.
% Participants wore the devices continuously, except when charging or engaging in activities that could damage the watch. 
% Every two hours, the watch app would activate and collect twelve minutes of photoplethysmography (PPG) signals, which were then utilized to extract heart rate and heart rate variability.
% The AWARE app monitored phone usage, while the mSavorUs app prompted participants to complete brief surveys multiple times daily. 
% This monitoring phase lasted around 8 weeks, and participants received compensation for their participation.

Participants meeting the inclusion criteria underwent a baseline assessment in a lab session and set up the devices under instruction.
During the monitoring phase for approximately 8 weeks, participants are asked to wear the devices continuously, except during charging or potentially damaging activities.
The watch app collected 12 minutes of photoplethysmogram (PPG) signals every 2 hours.
The AWARE app monitored phone usage, while the mSavorUs app prompted participants for brief surveys multiple times daily. 

\section{Method}

In this study, we devised a loneliness forecasting model on predicting loneliness seven days in advance. 
Given the complex nature of the multi-modal time series data, we employed machine learning models to perform feature extraction and classify the levels of loneliness. 
Our methodology encompasses the extraction of physiological, behavioral, social, and contextual features. 
We address challenges such as feature alignment, handling missing data, and the development of a classification model.

\subsection{Feature Extraction}
\subsubsection{Physiological Features}
We extract Heart Rate (HR) and Heart Rate Variability (HRV)-related features from the signal collected from smart rings and watches as the physiological features.
To this end, we extract the time-domain HRV characteristics (e.g. SDNN, RMSSD, AVNN), the frequency-domain HRV characteristics (e.g. LF, HF and LF / HF ratio) and the nonlinear HRV characteristics (e.g. SD1 and SD2) features~\cite{shaffer2017overview}.
In ubiquitous monitoring settings, signal collection with  PPG-based wearable devices can be affected by the presence of noise.
To overcome this issue, we implement the methods to extract HR and HRV features from raw collected PPG signals~\cite{feli2023energy,kazemi2022robust,wang2022ppg}. 
We first conduct a signal quality assessment to classify PPG signals as clean or noisy, then reconstruct short-term noisy segments via a generative adversarial network. 
Then the systolic peaks and inter-beat intervals are detected by a dilated Convolution Neural Network, from which HR and HRV-related features are extracted.

\subsubsection{Behavioral Features}
Behavioral features refer to the patterns and characteristics of subjects' smartphone usage behavior. 
They are extracted from participants' smartphone usage. 
We extract the smartphone usage features, including the number of battery charger plugins, screen off, screen on, screen locks, and screen unlocks.

\subsubsection{Social Features}
The term ``social features" pertains to the discernible patterns of social activity demonstrated by individuals through their smartphone usage. These features serve as valuable indicators that shed light on the nature of individuals' interactions and engagements within their respective social networks. Examples of such social features encompass the quantity of messages and notifications within various categories, as well as the duration and frequency of phone calls within specific time windows.

\subsubsection{Contextual Features}
Contextual features encompass the environmental information related to the subjects. 
These features provide insights into the surrounding context in which individuals operate using GPS localization. 
The contextual features include the variance of latitude, variance and mean of speed, number of places, duration at home, the mean and standard deviation of the duration outside, and the total travel distance within a given location time window.

\subsection{Feature Alignment}
The dataset contains features collected at varying resolutions, with some features recorded once daily (e.g., sleep-related) and others, such as PPG-related features, captured multiple times a day.
To align the features with label, we average the values within a designated time window preceding the corresponding loneliness levels.
 The selection of the optimal time window for each feature is based on the averaged feature's correlation with the target loneliness levels.
Subsequently, the resulting data records were compiled using the aligned features corresponding to the optimal window lengths and the self-reported loneliness level.

% \subsection{Data Aggregating}

% To facilitate the forecast of the loneliness category, we aggregate the physiological and behavioral features collected over a 14-day time window to a sole record. 
% However, during the data collection process, the number of data records within the time window may vary due to the inconsistent self-report frequency.
% To address this variability, we simplify the data by calculating the daily average for each feature. This approach ensures a fixed number of data points, corresponding to the number of days used (e.g., 14 days yield 14 data points), which enhances the training of our machine learning model.

\subsection{Missing Data}

In our dataset, missing data may occur due to participants forgetting to wear or charge their devices.
We discovered that using the average of all valid values for each participant exhibited a strong correlation with the loneliness label for most features. 
Therefore, we implement a single imputation method to address missing data across all features~\cite{donders2006gentle}.

\subsection{Classification Model}

% each participant's own median value, respectively. 
% By avoiding the use of the population median, we mitigate label imbalance and biased evaluations, ensuring a balanced distribution of labels.

% Each target loneliness category record was associated with a feature set over a 14-day time window that begins 3 weeks before the target. 
% As aforementioned, the features are paired with self-reported loneliness records.
% As a result, the number of feature records within the time window may vary due to the inconsistent self-report frequency.
% To address this variability, we distill the data by calculating the daily average for each feature. 
% This approach ensures a fixed number of data points, corresponding to the number of days used (e.g., 14 days yield 14 data points), which enhances the training of our machine learning model.
We define a binary classification task to forecast loneliness.
Each self-reported loneliness is associated with the aligned features within a 14-day time window, starting 3 weeks prior to the loneliness target. 
There is a one-week interval between the available features and the target. 
However, the number of aligned feature records within the time window may vary due to different self-report frequencies. 
To maintain a consistent count for each loneliness target, we aggregate the data by calculating the daily average for each aligned feature. 

% This aggregation process ensures that the number of features for each loneliness target remains consistent, corresponding to the number of time window length times the daily feature amount.
%For example, if the time window is set at 14 days, the resulting number of features will be 14 times the daily feature amount.
% This yielded a 14*208 dimensional feature set to train the model effectively.
% We then employ Random Forests for forecasting loneliness.
To forecast loneliness with the aggregated features, we employ Random Forests consisting of 400 trees with a maximum depth of 15.

\subsection{Model Explainability}

We explore the model explainability by utilizing path-dependent feature perturbation algorithms~\cite{NIPS2017_7062} to compute the Shapley values of the features.
%to gain insights into the feature importance of our trained personal models.
% Specifically, we employ path-dependent feature perturbation algorithms~\cite{NIPS2017_7062}, which are specifically designed for tree-based models. 
Based on the Shapley values for each feature, we obtain a quantitative measure of their contributions to the predictions.
These values allow us to rank the features based on their relative importance and assess their effect on the output loneliness category.
\section{Results}

\subsection{Loneliness Forecasting Performance}

% Our proposed method is evaluated on data from a two-month study involving 30 participants who completed several self-report assessments per day. 
% This results in a total of 6190 data points after preprocessing, and the development of 30 personal models. 
% We evaluated the trained model by comparing the estimated loneliness categories it produced with the corresponding category obtained from self-report, which served as the ground truth. 

% The performance is evaluated by averaging the results across all the personal models. 
% This allows us to assess the overall effectiveness of our approach in predicting loneliness levels.
% The overall performance of the loneliness forecasting model is shown in Table~\ref{tab:results}.
% Our model achieves an accuracy of 0.823, recall of 0.905, precision of 0.750, and f-1 of 0.820. The model was also evaluated using Cohen's kappa, which resulted in a value of 0.648, indicating substantial agreement between the predicted and true labels.

The models are trained and validated in a personalized manner on the data from a two-month study involving 29 participants.
Each participant's data were divided into a test set and a train set. The test set contains the most recent 50\% of the data in the monitoring. 
However, the training set includes the first 50\% data of the same person as well as data from all other participants. 
Personalized models were then trained and tested using these training-testing set pairs from each participant. 
This approach allows us to assess the model's performance by simulating real-world scenarios while maintaining the integrity of individual participant data.

% Our proposed method was evaluated using data from a two-month study involving 30 participants who completed multiple self-report assessments each day. 
After preprocessing, we obtained a total of 6212 data points, which were used to develop 29 personalized models to forecast loneliness 7 days ahead using the features from the past two weeks. 
To assess the performance of our approach, we average the results across all the personal models. 
%This allows us to evaluate the overall effectiveness of our method in predicting loneliness levels.
The overall performance is summarized in Table~\ref{tab:results}.
Our loneliness forecasting model achieves an accuracy of 0.823, recall of 0.905, precision of 0.750, F-1 score of 0.820, Cohen's kappa of 0.648.

\begin{table}[h]
    \vspace{-0.4cm}
    \caption{The overall performance on loneliness forecasting.}
    \label{tab:results}
    \vspace{-0.2cm}
    \centering
    \begin{tabular}{|c|c|c|c|c|}
    \hline
        Accuracy&Recall&Precision&F-1&Cohen's Kappa  \\
         \hline
         0.823 &0.905&0.750&0.820&0.648
 \\

\hline
       
    \end{tabular}

\end{table}

Table~\ref{tab:confusion} illustrates the confusion matrix of the loneliness detection models, derived from 3095 test samples.
This suggests that based on the physiological, behavioral, social, and contextual information, the model could forecast if the participant would feel lonely.
Note that the overall performance of our model is influenced by false negatives.
%This may provide the insight that in some ambiguous scenarios, feelings of loneliness may not necessarily cause variations in the physiological or behavioral outcomes.

\begin{table}[h]
\vspace{-0.4cm}
    \caption{Confusion matrix}
    \label{tab:confusion}
    \vspace{-0.2cm}
    \centering
    \begin{tabular}{|c|c|c|}
    \hline
        \backslashbox{Actual (label)}{Prediction}& Not Feel Lonely& Feel Lonely \\
         \hline
          Not Feel Lonely &1371&114
 \\

\hline
        Feel Lonely &412&1198\\
       \hline
    \end{tabular}
\vspace{-0.3cm}
\end{table}

% \begin{figure}
%     \centering
%     \includegraphics[width=0.35\textwidth]{confusion.png}
%     \caption{Confusion matrix illustrating the performance based on \textbf{N} test samples, provides an overview of the model's classification outcomes. The rows represent the actual ground truth labels, while the columns represent the predicted labels.}
%     \label{fig:confusion}
% \end{figure}

\subsection{Model Explainability}

\begin{figure}[h]
    \centering
     \vspace{-5mm}
    \includegraphics[width=0.5\textwidth]{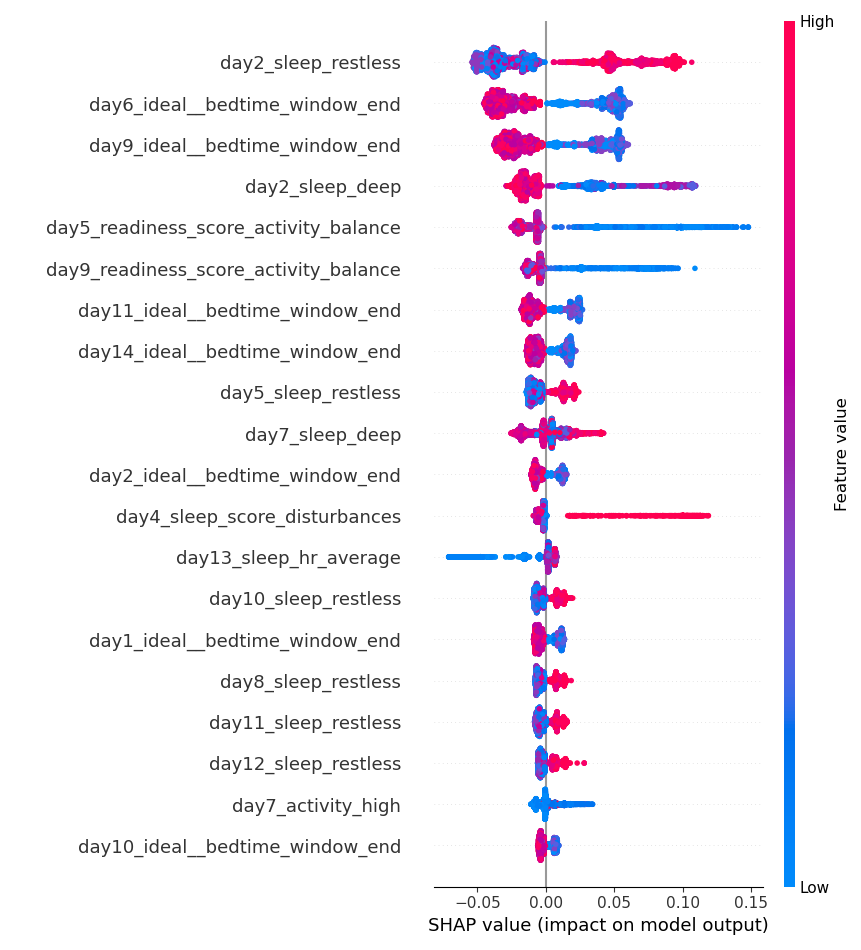}
    \vspace{-8mm}
    \caption{Feature Importance Using Shapley Values Beeswarm Plot}
    \label{fig:shap}
    \vspace{-5mm}
\end{figure}

Figure~\ref{fig:shap} presents the overall Shapley value across all test samples for the top 20 features.
The feature names are noted with the prefix 'day$x$' indicating the specific day within the time window from which each feature originates.
% Each dot in the plot represents an instance of a feature value, with its horizontal position indicating the effect of the feature on the predicted loneliness category. 
% Instances positioned more to the left suggest a higher likelihood of predicting 'Not Feeling Lonely,' while instances positioned more to the right indicate a higher likelihood of predicting 'Feeling Lonely'.
Each dot in the plot represents a feature value instance, with its horizontal position indicating its effect on the predicted loneliness category.
Dots positioned to the left suggest a higher probability of predicting ``Not Feel Lonely" while dots positioned to the right indicate a higher probability of predicting ``Feel Lonely".

% The most influential feature is the sleep restless score, which represents the percentage of time during sleep when the participant was moving.
% A higher sleep restless score is associated with a higher likelihood of loneliness.

We observe that higher sleep restless scores are strongly associated with increased loneliness likelihood, making it the most influential feature.
% The end time of the ideal bedtime window is another informative feature that reflects the participant's potential to go to sleep at the optimal timing. 
% A later end time suggests a greater ease of falling asleep within the window and sleeping at ideal timing leads to a lower likelihood of feeling lonely.
% Another group of significant features is the activity balance score, where a lower value indicates a lower level of activity compared to normal.
% In addition, less activity is potentially associated with a higher likelihood of feeling lonely.
% Later end times of the ideal bedtime window are indicative of a higher likelihood of falling asleep within the optimal timing, resulting in reduced loneliness.
% Delayed end time of the ideal bedtime window increases the probability of falling asleep within the optimal timing, promoting adherence to the ideal sleep schedule and potentially reducing feelings of loneliness.
Lower activity balance scores also contribute significantly, implying that lower activity levels are associated with an increased likelihood of feeling lonely.
Clinicians can address the fact that poor sleep and low levels of activity are connected to loneliness, by assisting clients in improving these aspects of their daily lives. Even though it might not be the initial line of treatment that clinicians consider for loneliness, it could hold substantial importance in addressing the issue.

\section{Conclusion}

This paper proposed a machine learning method for forecasting loneliness based on physiological, behavioral, and
contextual information collected from smart rings, watches, and smartphones
Our developed model achieved an accuracy of 0.823 and an F-1 score of 0.820 in forecasting loneliness levels 7 days ahead. 
We incorporated Shapley values as explainability methods to gain insights into the relationship between the features and loneliness. 
By extending beyond existing studies that focus on prompt loneliness detection, our approach offered the potential for clinical implementation in the early identification and addressing of loneliness among at-risk populations. 
Further investigation into the utilization of deep learning models for predicting loneliness has the potential to improve performance by automatically extracting more robust features. Moreover, deep learning holds promise in forecasting loneliness in several weeks.
% The curated comprehensive dataset, encompassing wearable device data, behavioral information, and self-reported loneliness levels, paved the way for further research to develop impactful interventions for loneliness.

\bibliographystyle{ieeetr}
\bibliography{references}

\end{document}